\documentclass[aps,preprint,floatfix,pre]{revtex4-1}
\usepackage{graphicx}

\begin{document}
\title{Gyrokinetic Studies of Microinstabilities in the RFP}
\author{D. Carmody$^1$}
\email{dcarmody@wisc.edu}
\affiliation{$^1$University of Wisconsin-Madison, Madison, Wisconsin 53706, USA}
\author{M.J. Pueschel$^1$}
\affiliation{$^1$University of Wisconsin-Madison, Madison, Wisconsin 53706, USA}
\author{P.W. Terry$^1$}
\affiliation{$^1$University of Wisconsin-Madison, Madison, Wisconsin 53706, USA}
\date{\today}



\begin{abstract}
An analytic equilibrium, the Toroidal Bessel Function Model, is used in conjunction with the gyrokinetic code GYRO to investigate the nature of microinstabilities in a reversed field pinch (RFP) plasma. The effect of the normalized electron plasma pressure ($\beta$) on the characteristics of the microinstabilities is studied. A transition between an ion temperature gradient (ITG) driven mode and a microtearing mode as the dominant instability is found to occur at a $\beta$ value of approximately 4.5\%. Suppression of the ITG mode occurs as in the tokamak, through coupling to shear Alfv\'en waves, with a critical $\beta$ for stability higher than its tokamak equivalent due to a shorter parallel connection length. There is a steep dependence of the microtearing growth rate on temperature gradient suggesting high profile stiffness. There is evidence for a collisionless microtearing mode. The properties of this mode are investigated, and it is found that curvature drift plays an important role in the instability. 
\end{abstract}

\maketitle


\section{Introduction}

Confinement in the reversed field pinch (RFP) is dominated by global tearing modes under ordinary circumstances.  Consequently, microscale drift-type fluctuations have received far less attention in the RFP than they have in the tokamak.  However, operational modes such as pulsed poloidal current drive \cite{Sarff1994}, which applies an external force to flatten the current gradient,  and the quasi-single-helicity state \cite{Escande2000}, whose quasi-spontaneous formation leads to a suppression of multiple helicity tearing modes, have gained success in reducing global tearing mode activity.  In this regime, confinement time can improve significantly, and equilibrium gradients can steepen \cite{Chapman2002,Kim2012}.  This produces conditions in which microscale turbulence might emerge as a factor in confinement, just as it does in the tokamak.  The experimental evidence for fluctuations that are independent of global tearing modes is still fragmentary but growing.  The tearing modes couple to a broad cascade that delivers magnetic energy to small scales and simultaneously brings kinetic energy closer to equiparition as the wavenumber increases \cite{Thuecks2012}, in keeping with MHD.  However, at smaller scales, the kinetic energy eventually becomes greater than the magnetic energy.  In this regime there are notable changes in the coherence and cross phase between magnetic field and density, likely associated with a different type of fluctuation.  The observed radial structure indicates a standing wave pattern consistent with collisional shear Alfv\'{e}n waves or microtearing modes \cite{Ren2011}.  There is also evidence that the temperature gradient in the edge is close to a critical gradient for temperature gradient driven microturbulence \cite{Sarffa}.

After years of some uncertainty about what type of microinstability is most compatible with the high shear and low safety factor conditions of the RFP, numerical solutions of gyrokinetic models in RFP geometry have shown that many modes familiar from the tokamak may also arise in the RFP \cite{Predebon2010a, Predebon2010, Tangri2011}. For example, ion temperature gradient (ITG) driven modes can dominate instability at zero $\beta$, above its critical temperature gradient, albeit with a somewhat different character from their tokamak counterpart.  ITG is a curvature drift resonant mode, and in the RFP it is the poloidal curvature that dominates.  With bad curvature for all poloidal angles, the mode has a longer poloidal extent instead of a ballooning character \cite{Predebon2010}.  The scale length of the critical temperature gradient for instability goes like the minor radius $a$ instead of the major radius $R_0$.

Analysis of instabilities that only arise for zero or very low values of $\beta$ makes little sense for the RFP, where $\beta$ values of 10\% are not unusual.  When increasing $\beta$, it is reasonable to expect that the ITG mode is eventually stabilized -- it is thus helpful to know at what $\beta^{\mathrm{ITG}}_{\mathrm{crit}}$ stabilization occurs and how $\beta^{\mathrm{ITG}}_{\mathrm{crit}}$ scales with  various geometric and equilibrium parameters.  It is also important to determine whether a new instability emerges at higher $\beta$, and if so, at what critical $\beta$ the instability arises and what characteristics apply to it.  In the tokamak, the kinetic ballooning mode becomes unstable as $\beta$ is increased beyond a certain threshold.  In the RFP, it would not be surprising for a microtearing (MT) mode \cite{Gladd1980,Drake1980,Drake1977,Connor1990} to become unstable for $\beta$ values above the low-$\beta$ ITG regime, provided the electron temperature gradient that drives the MT mode is comparable to the ion temperature gradient for ITG instability.  
The MT mode is a natural fluctuation for the RFP, if only because it is the small-scale extension of global tearing modes, which dominate confinement in ordinary discharges.  The MT mode is not a current driven mode: current driven tearing modes require that a current gradient that varies on the scale of the minor radius $a$ be larger than the flux discontinuity at the resistive layer, which goes like $-2k_\theta$, ensuring that $\Delta ' > 0$, where $\Delta'$ is the standard parameter of the MHD tearing mode.  With the binormal wavenumber $k_\theta$ large, microtearing modes have $\Delta ' < 0$.  However, they can be driven by electron temperature gradients just like other drift modes, including trapped electron modes and electron temperature gradient driven modes.

We show in this paper that under certain conditions, including high temperature gradients and weak density gradients, MT indeed supplants the ITG instability as the dominant mode above a critical $\beta$ value around 5\% - 6\%.   In this transition the dominance of ITG at low $\beta$ is assured by having an ion temperature gradient above the threshold $a/L_{Ti} \approx 3$, where $1/L_{Ts}=\nabla T_s / T_s$ is the temperature gradient scale length and $s$ indicates the particle species.  The dominance of microtearing at higher $\beta$ is assured by having the electron temperature above a similar threshold ($a/L_{Te} \approx 3$).  We show that the coupling to Alfv\'{e}n waves that stabilizes the ITG mode as $\beta$ increases does so at a higher $\beta$ in the RFP than in the tokamak because of the shorter parallel connection length associated with the large poloidal field.  The MT mode is identified on the basis of a mode structure that shows canonical tearing parity in the electromagnetic fields and a frequency in the electron direction.  We examine scaling properties of the MT growth rate, including its threshold behavior, its scalings with the temperature ratio, and equilibrium quantities like the pinch parameter $\Theta$.  Growth rate scalings with respect to both $r$ and $\Theta$ relate to magnetic shear scaling, which in the RFP is not an independent equilibrium parameter.  This work follows a number of recent studies that suggest the microtearing mode may be important in both the tokamak and the RFP \cite{Applegate2007,Doerk2011,Hatch2012,Guttenfelder2011, Predebon2010a}.  Here our analysis is specific to equilibria consistent with the Madison Symmetric Torus \cite{Prager1990}.  Moreover, in adapting gyrokinetic codes originally developed for the tokamak, we are careful to capture all of the effects arising from an equilibrium magnetic field with comparable poloidal and toroidal field components.  This can be done using the toroidal Bessel function model in the GYRO code \cite{Candy2003a}, with RFP-appropriate representations of the curvature drift and parallel derivative \cite{Tangri2011}.  We demonstrate here through benchmarking studies that quantitatively idential results can be obtained with corresponding modifications to the circular equilibrium model \cite{Lapillonne2009} of the {\sc Gene} code \cite{Jenko2000,GENEpage}. 

An interesting aspect in recent gyrokinetic work relating to the MT mode is the observation of instability in low-collisionality regimes.  This is observed for simulations with both tokamak \cite{Applegate2007,Doerk2011, Guttenfelder2011} and RFP geometries \cite{Predebon2010a}.  Theoretically, a series of papers specific to the tokamak in their approximations have collectively pointed to the conclusion that the microtearing mode is very sensitive to collisionality $\nu$, and should become stable for small collisionality \cite{Connor1990, Catto1980, Drake1977, Hassam1980a}.  One design of the current study has been to probe this mismatch between these theoretical predictions and the results of the aforementioned gyrokinetic simulations.  A collisionality scan shows instability at low $\nu$, but behavior with other parameters suggests that there may be two branches of the instability, one for low collisionality and one for higher values more compatible with theoretical predictions.  We specifically study the possibility, first suggested for RFP tearing modes in Ref.~\cite{Finn1986}, of an RFP MT branch that is enabled by the large electron curvature drift of the RFP.  Artificial variation of the electron curvature drift strength shows that the growth rate diminishes toward zero when the curvature drift falls outside a certain range of values.
 
This paper is organized as follows: First, the gyrokinetic model is introduced in Sec.~\ref{model}. Included in this section is a discussion of the geometric features of the RFP and how they are dealt with in GYRO, the gyrokinetic code used in the majority of these results. Then, in Sec.~\ref{beta}, we show results from a $\beta$ scan for a certain set of RFP parameters and discuss the instabilities seen at low and high $\beta$. This is followed by a derivation of finite-$\beta$ suppression of ITG in Sec.~\ref{ITG}. Next, Sec.~\ref{micro} contains a detailed characterization of the RFP microtearing mode, including an investigation of growth rate dependencies on parameters such as collisionality, temperature gradient, and curvature. All of these results are summarized in Sec.~\ref{conc}.

\section{Gyrokinetic Modeling \label{model}}
A comprehensive method by which to investigate microinstabilities in toroidal plasma devices is the gyrokinetic framework, in which the fast gyromotion has been removed from the kinetic equations through a gyrophase averaging procedure \cite{Brizard2007}, reducing the problem from the original 6D phase space to a more computationally tractable 5D.

In this work we present results from the initial-value gyrokinetic code GYRO \cite{Candy2003a}. This code provides a numerical solution of the gyrokinetic Vlasov-Maxwell system of equations. We here restrict ourselves to linear simulations in a radially local (flux-tube) domain, in which the main quantities of interest are the growth rates and frequencies of the most unstable mode at a particular wavenumber $k_\theta\rho_s$, where $\rho_s=(T_e/m_i)^{1/2}m_i/eB$ is the ion sound gyroradius. This linear analysis provides an initial characterization of microinstabilities in the Madison Symmetric Torus (MST) \cite{Dexter1990}. An expanded nonlinear analysis is currently in progress, and this nonlinear work, in conjunction with measurements from up and coming diagnostics on MST, will provide a unique opportunity for validation of the gyrokinetic results.

In order to properly capture the effects of the magnetic geometry of the RFP, a number of modifications have been made to the code. These changes concern both the magnetic equilibrium used and the expression of certain geometric operators. A more detailed explanation of the changes and their effect on linear growth rates is given in Ref.~\cite{Tangri2011}, and only an outline will be given here.

The fundamental change that has been made is the inclusion of the Toroidal Bessel Function Model (TBFM) as the equilibrium model used in GYRO. The standard equilibrium model in the RFP is the Bessel Function Model (BFM) [$B_\theta = B_0 J_1(\lambda r), B_\phi = B_0 J_0(\lambda r)$] which is the solution of the force free condition $\nabla \times {\bf B} = \lambda {\bf B}$ in cylindrical geometry. The TBFM is an extension of the BFM to toroidal geometry, and since it is derived from the Grad-Shafranov equation, it is able to incorporate finite pressure effects. Flux surfaces are assumed to be circular, and to lowest order in $\beta$ the magnetic fields are given as:
\begin{equation}
B_\theta = \frac{B_0 J_1(2 \Theta r/a)}{1+r\cos\theta/R_0},\; B_\phi = \frac{B_0 J_0(2 \Theta r/a)}{1+r\cos\theta/R_0}
\label{TBFM}
\end{equation}
where $\Theta=\langle B_\theta\rangle^\mathrm{wall} / \langle B_\phi \rangle^\mathrm{vol}$ is the RFP pinch parameter. In these simulations, $\Theta$ is an important input parameter, from which the safety factor $q_0$ and shear $\hat s$ are calculated self-consistently. The magnetic geometry in the local simulations is thus completely determined by $\Theta$ and the normalized radius $r_0/a$ of the simulation domain.

In addition to the background magnetic equilibrium, certain of the gyrokinetic operators are generalized to the RFP geometry. These generalizations stem from characteristics of the RFP equilibrium that differentiate it from the tokamak. Primarily, the poloidal magnetic field in the RFP is much stronger relative to the toroidal field than in the tokamak, and in the case where the normalized radius of the simulation domain is $r_0/a \sim 0.5$ (a typical region of interest in simulations), they are roughly the same order of magnitude. Under these circumstances, the common tokamak approximation $B\sim B_\phi$ cannot be made, and the more general form $B = B_\phi(1+(\epsilon_t/q_0)^2)^{1/2}$ must be used, where $B_\phi$ is the toroidal magnetic field and $\epsilon_t=r/R_0$ is the inverse aspect ratio. This difference affects operators such as the curvature drift frequency and parallel transit operators, the latter of which in its general form will pick up an additional factor,
\begin{equation}
{\bf b} \cdot \nabla \sim k_\parallel = \frac{1}{\sqrt{1+(\epsilon_t/q_0)^2}}\frac{1}{q_0R_0},
\end{equation}
Besides being incorporated into the code, these geometric modifications must also be taken into account in analytic estimates whenever generalizing a tokamak analysis to an RFP environment. The next section provides an example of this in the context of finite-$\beta$ suppression of ITG.

The results presented below were performed using a set of parameters chosen to represent the experimental conditions in MST. This device has a major radius of $R_0 = 1.5\,\mathrm{m}$ and a minor radius of $a = 0.5\,\mathrm{m}$, yielding an inverse aspect ratio of $\epsilon_t=1/3$. We look at the radial location $r_0/a=0.5$, with, unless otherwise stated, the other parameters being: $q_0=0.186, \hat s=-0.716, \Theta=1.35, a/L_n = 0.58, a/L_T = 5.0, T_i/T_e = 0.4, \nu (a/c_s)=0$. The collisional frequency $\nu$, in particular, plays an important role in the dynamics of the MT mode, which will be discussed in Sec.~\ref{micro}.

Now we look at the effect of $\beta$ in determining the nature of the dominant linear instability.

\section{Beta scan \label{beta}}

\begin{figure}[p]
\begin{center}
\includegraphics[angle=0,scale=0.59]{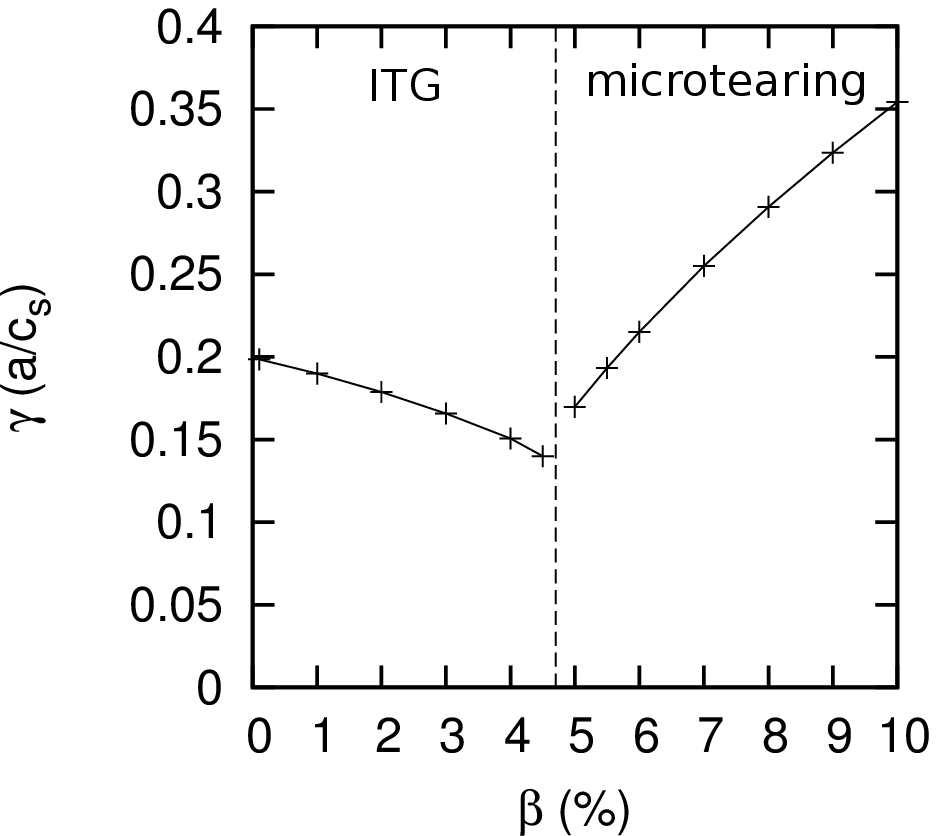}
\includegraphics[angle=0,scale=0.59]{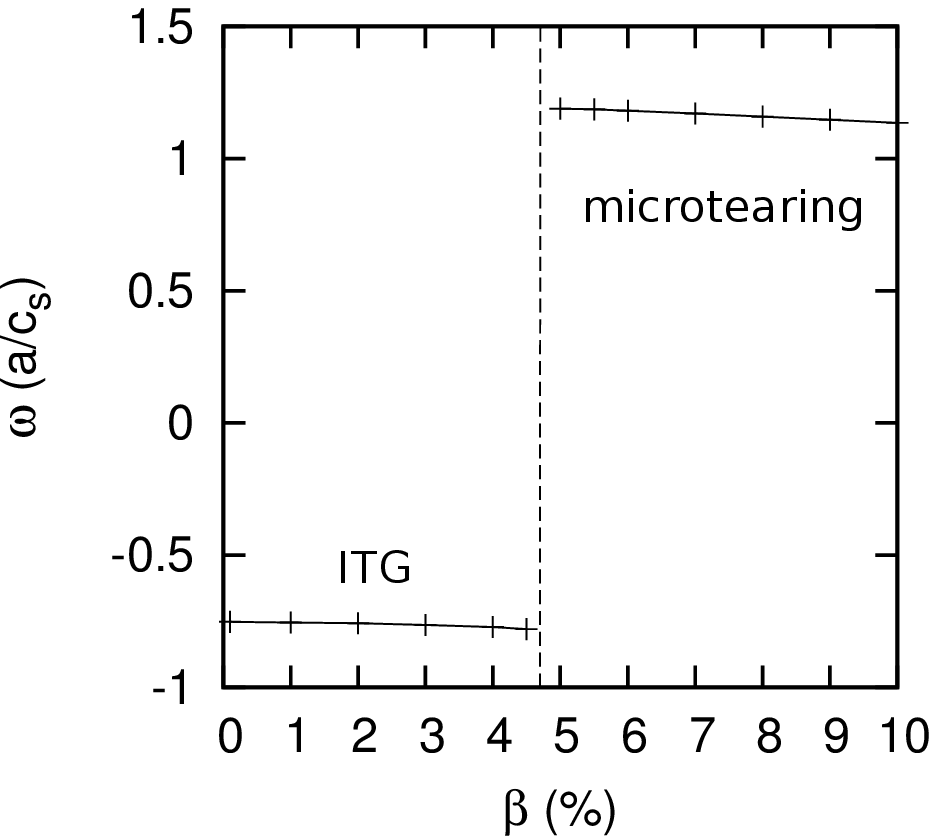}
\end{center}
\caption{Growth rate and mode frequency plotted as a function of $\beta$ for $k_\theta\rho_s = 0.372$. A transition of the dominant mode from ITG to MT occurs at $\beta\approx5\%$}
\label{betascan}
\end{figure}

A scan over $\beta$ from $0$ to $10\%$ was performed to determine the variety of modes that might be dominant across this range. The results can be seen in Fig.~\ref{betascan}. It should be noted that for this scan, the pinch parameter is kept constant and does not vary self-consistently with $\beta$. At low $\beta$, the dominant instability is identified as an electrostatic ITG mode. As $\beta$ increases this mode is suppressed and eventually overtaken by a MT mode at a $\beta$ of approximately $4.5\%$. The transition can be seen most clearly in the real frequency plot. Here, the frequency of the ITG mode is in the ion direction (negative sign in this convention) and that of the MT mode is in the electron direction (positive sign).

\begin{figure}[p]
\begin{center}
\includegraphics[angle=0,scale=0.59]{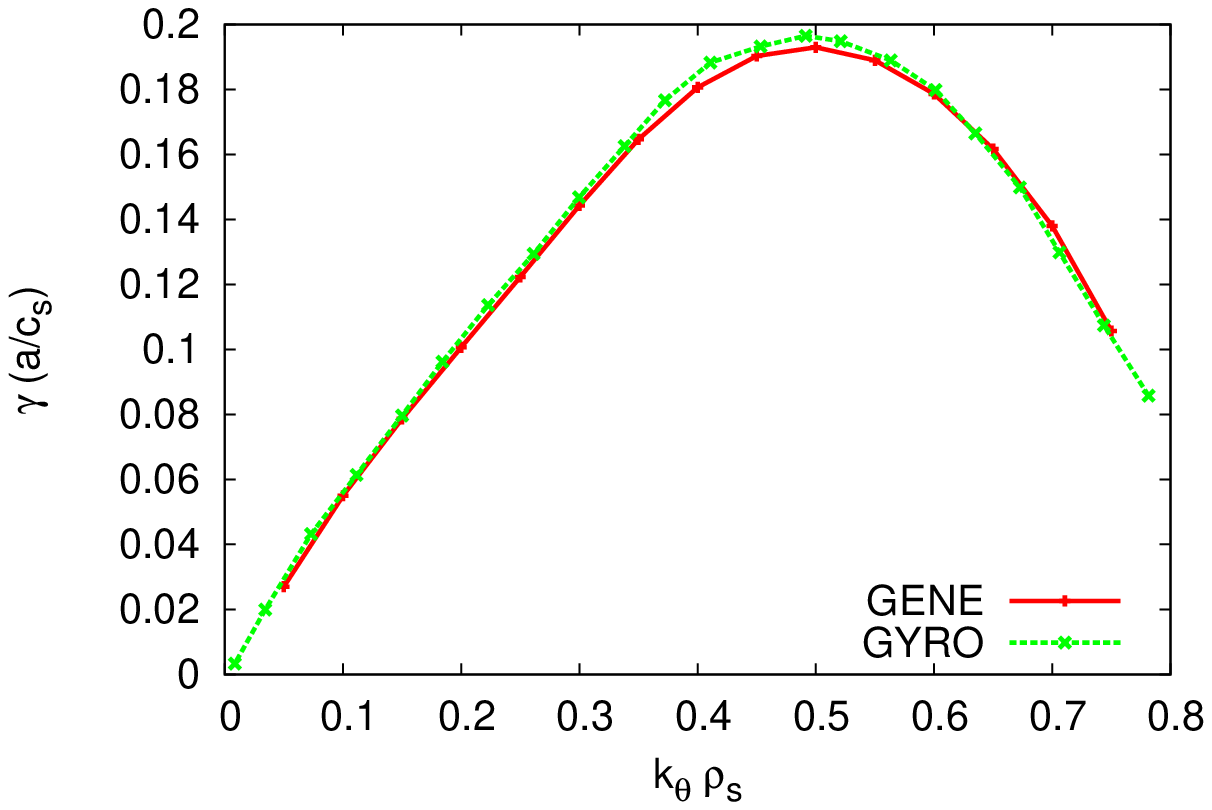}
\includegraphics[angle=0,scale=0.59]{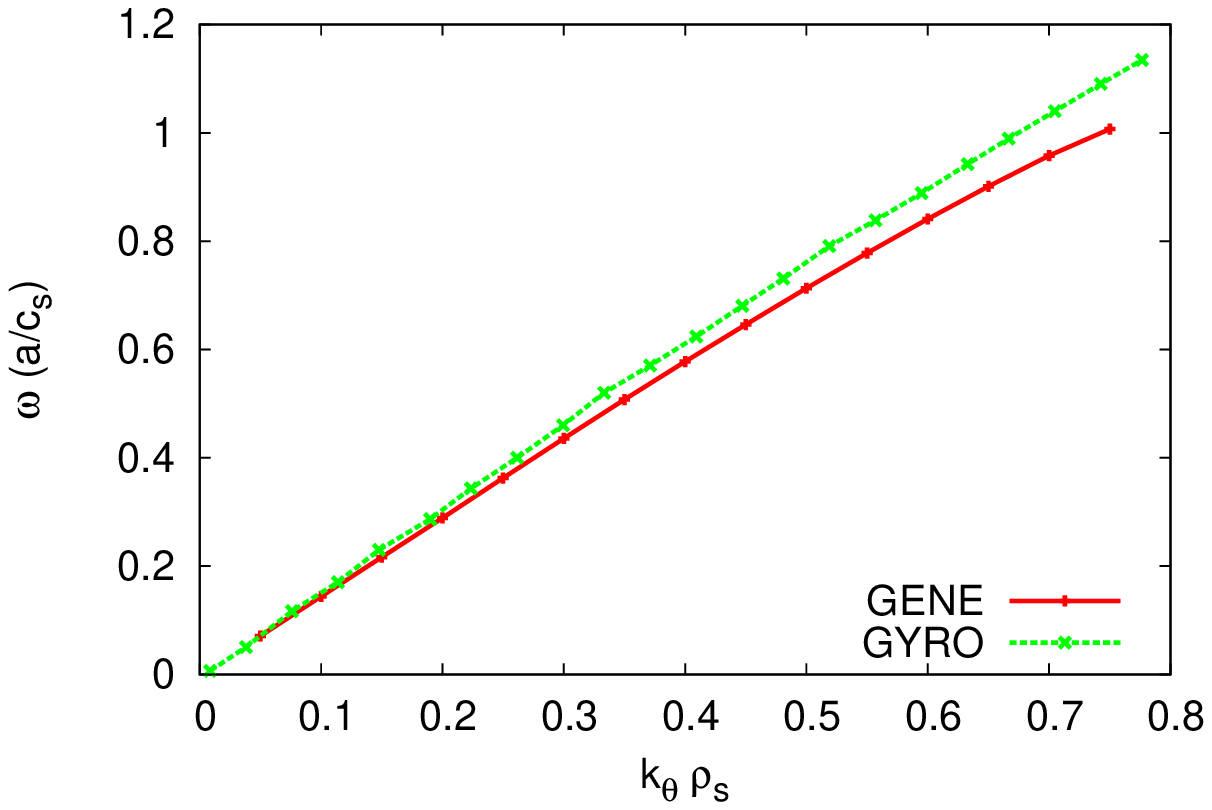}
\end{center}
\caption{(Color online) A comparison of {\sc Gene} (red $+$'s) and GYRO (green $\mathrm{x}$'s) results for $\beta=0$ in the RFP. Other parameters are given in the text. The two codes show good agreement. The difference in real frequency at high $k_\theta$ may be a results of incomplete convergence.}
\label{benchmark}
\end{figure}

These results have been confirmed with the circular equilibrium in the {\sc Gene} code for the $\beta=0$ ITG regime (see Fig.~\ref{benchmark}). Since gyrokinetic simulations are a relatively new topic of study for the RFP, it is important to compare results from separate codes with slightly different handling of geometric quantities. In order to attain agreement between the codes it was necessary to generalize {\sc Gene}'s circular equilibrium by including two additional terms to $\mathrm{d}B/\mathrm{d}r$: a term proportional to $\hat s$, and a term accounting for the $r$ dependence of $B_\phi$.

\begin{figure}[p]
\begin{center}
\includegraphics[angle=0,scale=0.6]{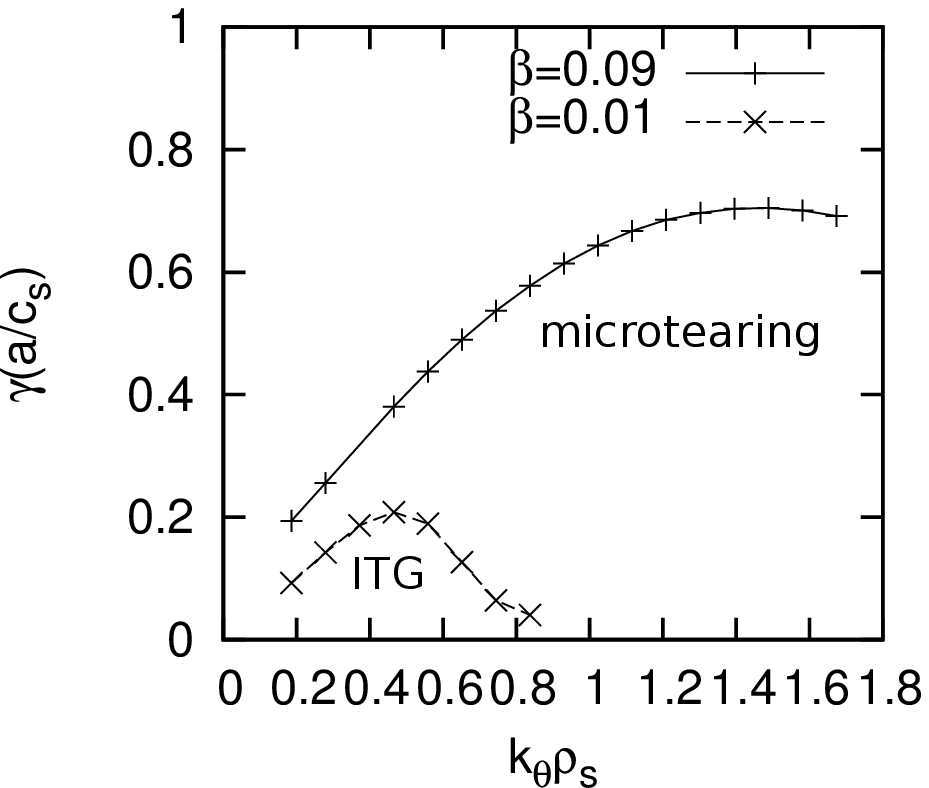}
\includegraphics[angle=0,scale=0.6]{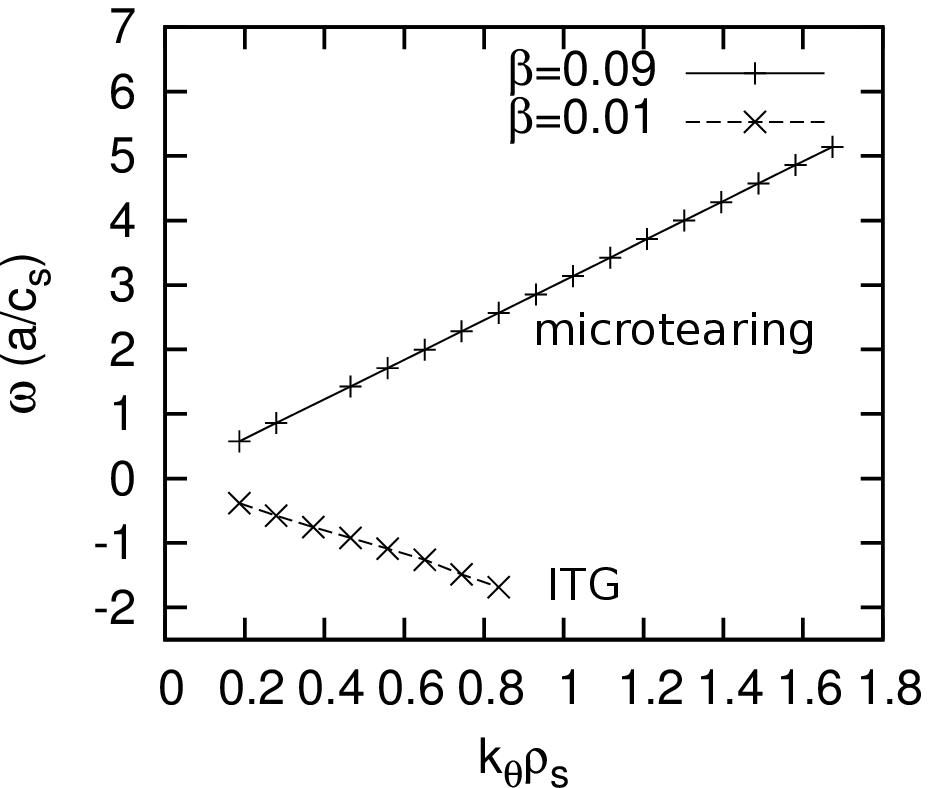}
\end{center}
\caption{Growth rate and frequency as a function of wavenumber $k_\theta\rho_s$ for two different values of $\beta$. ITG is seen to be dominant at $\beta_e=1\%$, microtearing at $\beta_e=9\%$}.
\label{b9b1}
\end{figure}

One interesting feature of the low-$\beta$ versus high-$\beta$ instabilities is the range of scales at which these modes are unstable. This difference can be seen in Fig.~\ref{b9b1}. At low $\beta$, the ITG mode ranges from $k_\theta \rho_s=0.1$ to $k_\theta \rho_s=0.9$, achieving a peak near $0.5$. Initially, the MT mode arises at these same scales, with a peak at roughly the same value, though with a slightly broader range. As $\beta$ increases, however, the peak of the microtearing mode shifts to higher $k_\theta$. This can be seen in Fig.~\ref{b9b1}, where it is observed that at a $\beta=9\%$ the MT mode not only reaches higher growth rates, but does so over a much larger range of scales, peaking at a value of $k_\theta\rho_s \approx1.5$.

The range of $\beta$ simulated covers different modes of operation of MST. A standard discharge will have values of $\beta\sim 4-5\%$, a range that means ITG and microtearing may both be equally strong. Improved confinement pulsed poloidal current drive (PPCD) discharges, on the other hand, may achieve $\beta$ values of $9\%$ or higher \cite{Chapman2009}, in which case microtearing may be the dominant mode. Runs with $\beta = 9\%$ will be the focus of Sec.~\ref{micro}. In the next section, we discuss the observed suppression of ITG with increasing $\beta$.

\subsection{ITG beta suppression \label{ITG}}

\begin{figure}[p]
\begin{center}
\includegraphics[angle=0,scale=0.6]{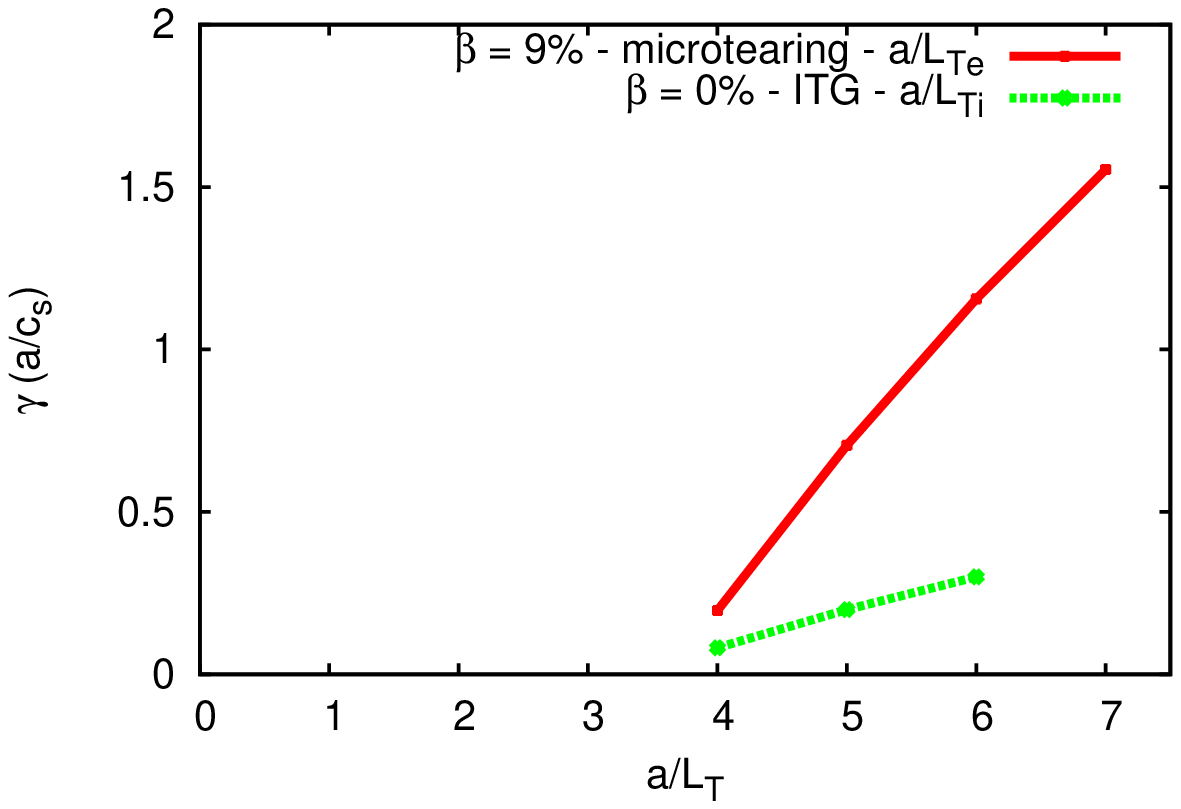}
\caption{(Color online) Growth rate plotted against temperature gradient for $k_\theta\rho_s=0.372$ in the case of MT (red squares) and ITG (green circles). Both instabilities have a threshold around $a/L_{T} \approx 3.5-4.0$, for their respective driving gradients.}
\end{center}
\label{ltscan}
\end{figure}

Finite-$\beta$ suppression of ITG (a linear effect which may be amplified nonlinearly \cite{Pueschel2010a}) has been a topic of study in the context of tokamaks \cite{Hirose2000}, and that analysis will be applied here to ITG in the RFP. In the RFP, ITG growth rates can still be quite strong at values of $\beta$ where tokamak ITG is typically stable \cite{Candy2003a}, as can be seen in Fig.~\ref{betascan}. In fact, ITG remains unstable past $\beta =5\%$ (though subdominant to microtearing) and may not stabilize until $\beta \sim 10\%$. 

Our analysis of the finite-$\beta$ suppression of the ITG mode follows that of Hirose \cite{Hirose2000}, making the appropriate modifications for the RFP geometry (see Sec.~\ref{model}). As discussed, these modifications are due to the different strengths and scale lengths of the magnetic field. In the tokamak, the scale length of magnetic field variation is proportional to the major radius, $ 1/L_B = \nabla B / B \sim 1 / R_0 $, while in the RFP, the appropriate scale is the minor radius, $ \nabla B / B \sim 1 / a $. As was mentioned above, the parallel derivative term also needs to be modified to account for the equivalent strengths of the poloidal and toroidal fields. Therefore, the modified terms will take the forms $ k_\parallel = 1/(q_0R_0(1+(\epsilon_t/q_0)^2)^{1/2}) $ and the curvature drift $ \omega_{De} = 2cT_e(\nabla {\bf B} \times {\bf B}) \cdot {\bf k} /eB^3 \sim 1/L_B$, where $c$ is the speed of light and $e$ is the fundamental charge.

The following discussion is based on a fluid analysis of the ITG instability. Before including the finite-$\beta$ effects, the ion and electron densities are found to be, respectively,

\begin{equation}
n_i = \frac{(\omega+5\omega_{Di}/3)(\omega_{*e}-\omega_{De}) - (\eta_i-2/3)\omega_{*i}\omega_{De}}{(\omega + 5\omega_{Di}/3)^2 - 10\omega_{Di}^2/9},
\end{equation}
\begin{equation}
n_e = \frac{e\Phi}{T_e}n_0,
\end{equation}
where $\omega_{*e} = cT_e(\nabla {\ln n_0} \times {\bf B}) \cdot {\bf k} / eB^2 \sim 1/L_n$ and $\eta_i = d\ln T_i / d\ln n_0$.

The expression for the electron density is modified by the consideration of finite-$\beta$ effects. This is done by taking into account perpendicular magnetic field perturbations, or, equivalently, perturbations to the parallel magnetic vector potential: ${\bf B}_\perp = \nabla \times {\bf A}_\parallel$.
Including such perturbations in the parallel momentum balance of electrons will result in the electron density taking the form

\begin{equation}
n_e = \left( \Phi - \frac{\omega - \omega_{*e}}{ck_\parallel}A_\parallel \right) \frac{en_0}{T_e},
\end{equation}

Using Amp\`ere's law and the quasineutrality condition, we arrive at the following relation,

\begin{equation}
A_\parallel \left(1 - \frac{\beta}{k^2_\parallel L_n L_B}\left[2\frac{\varepsilon_n}{\tau^2} + \frac{1}{\tau}\left(1 + 2\varepsilon_n\right) + 1 + \eta_e\right]\right) = \frac{\omega_{*e}}{ck_\parallel}\frac{k^2_{De}}{k^2_\perp}\left(1 + 2\frac{\varepsilon_n}{\tau}\right)\Phi,
\end{equation}
where $\tau = T_e/T_i$, $k_{De} = (4\pi n_0 e^2 / T_e)^{1/2}$, $\eta_e = d\ln T_e / d\ln n_0$, $k_\perp={\bf k}\cdot{\bf B}_\perp\approx k_\theta$, and $\varepsilon_n = L_n/L_B$. Then, as in Ref.~\cite{Hirose2000}, we are able to derive the stability condition, with certain terms adjusted to account for the RFP generalizations,
\begin{equation}
\beta \ge \frac{\varepsilon_n \epsilon_t^2 \tau^2}{(1+(\epsilon_t/q_0)^2)q_0^2[(\tau+2\varepsilon_n)(\tau + 1) + \tau^2\eta_e]},
\end{equation}
Choosing parameters similar to those used for the growth rates plotted in Fig.~\ref{betascan} ($\tau=2.5,\varepsilon_n=1/.58,\eta_e=8.6,q_0=1/5,\epsilon_t=1/3$), the above expression yields a critical $\beta$ of approximately $10.6\%$. This agrees well with an estimate of $\beta^{\mathrm{ITG}}_{\mathrm{crit}}$ from Fig.~\ref{betascan} based on extrapolation of the ITG growth rate for $\beta > 5\%$.

The criterion given above is similar to that of Ref. \cite{Hirose2000}, but for RFP parameters it yields a higher critical $\beta$ than is seen in tokamaks. This is due primarily to the smaller $q_0$ and shorter connection length that is a result of the equivalent strengths of the poloidal and toroidal fields in the RFP. 

We now turn our attention to the instability observed at higher $\beta$, the microtearing mode.

\subsection{Microtearing Modes\label{micro}}

\begin{figure}[p]
\begin{center}
\includegraphics[angle=0,scale=0.6]{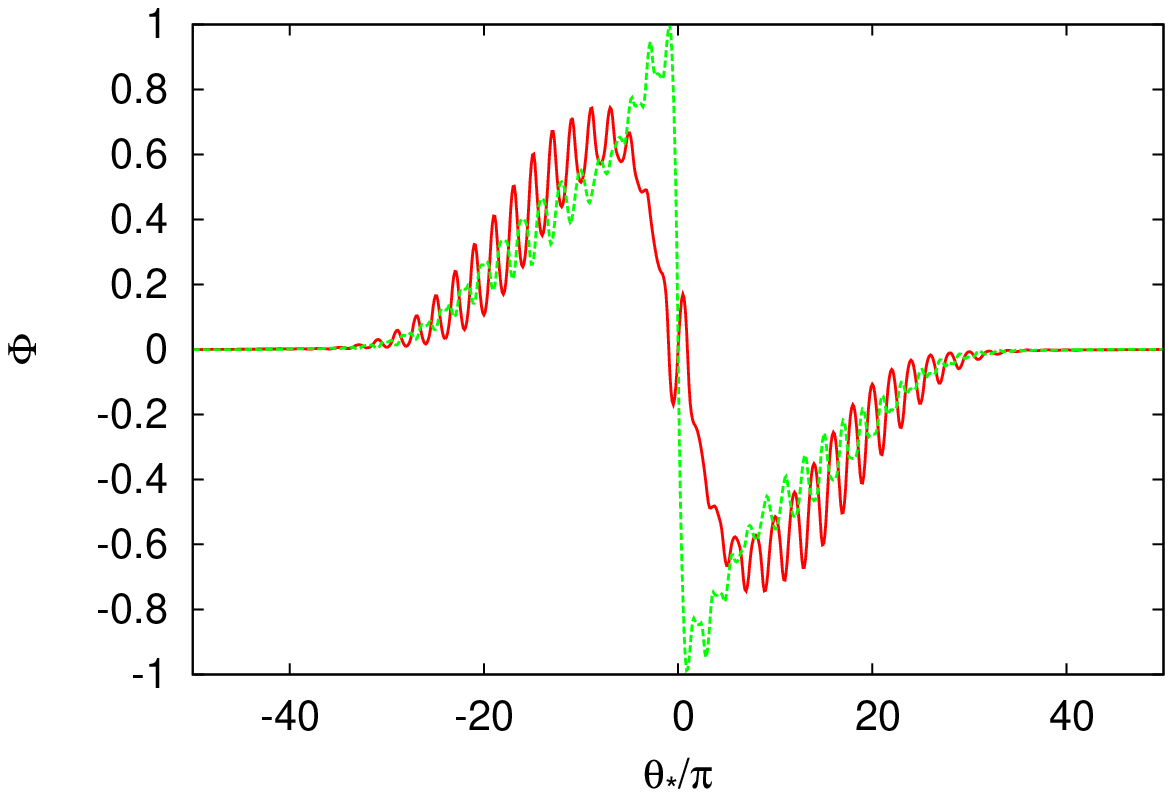}
\includegraphics[angle=0,scale=0.6]{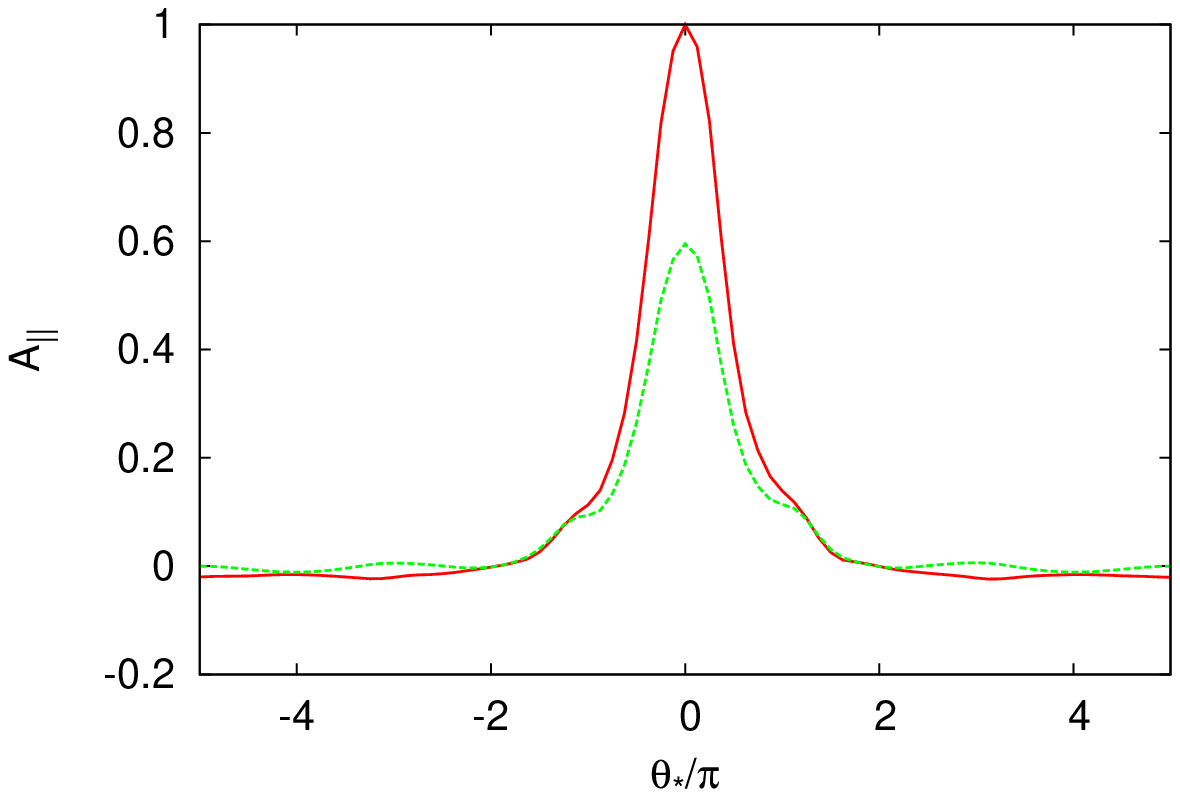}
\end{center}
\caption{(Color online) Eigenmode structure for the MT mode in electrostatic potential $\Phi$ and magnetic potential $A_\parallel$ with both real (green dashed curve) and imaginary (red solid curve) components. The fields are plotted against the magnetic-field following ballooning angle $\theta_*$. This mode displays tearing parity, which is recognized as even parity in $A_\parallel$ and odd parity in $\Phi$.}
\label{eigstructure}
\end{figure}

The dominant instability at higher values of $\beta$ is identified as a MT mode. These modes are characterized by having tearing parity in the parallel direction. This appears as odd parity in the electrostatic potential $\Phi$ and even parity in the  magnetic potential $A_\parallel$, as seen in the eigenmode structure in Fig.~\ref{eigstructure}. This structure, including the small amplitude features occurring every $2\pi$ (and attributed to poloidal variation) is similar to that seen in other devices, including MAST \cite{Applegate2007} and RFX-mod \cite{Predebon2010a}. Several parameters scans were performed in order to better characterize the observed MT mode. These scans were performed at a $\beta$ value of $9\%$, chosen to lie in the range for PPCD discharges in MST.

\begin{figure}[p]
\begin{center}
\includegraphics[angle=0,scale=0.6]{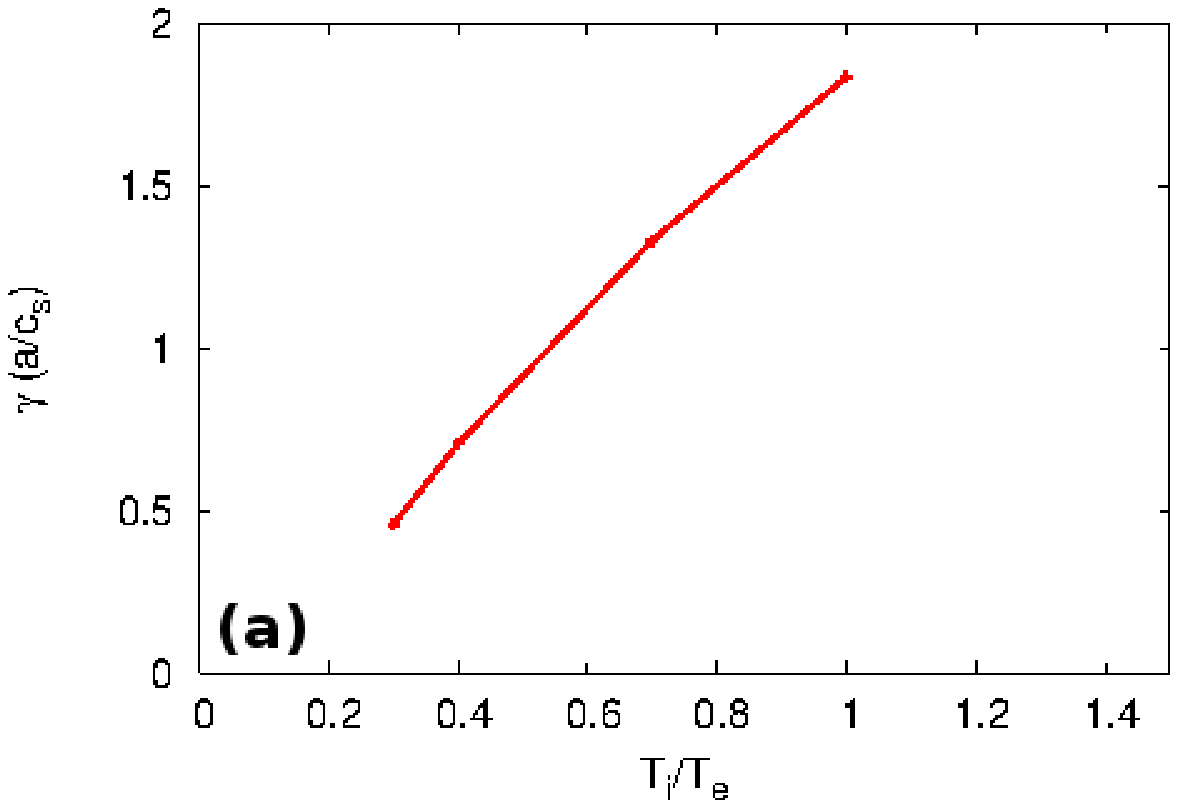}
\includegraphics[angle=0,scale=0.6]{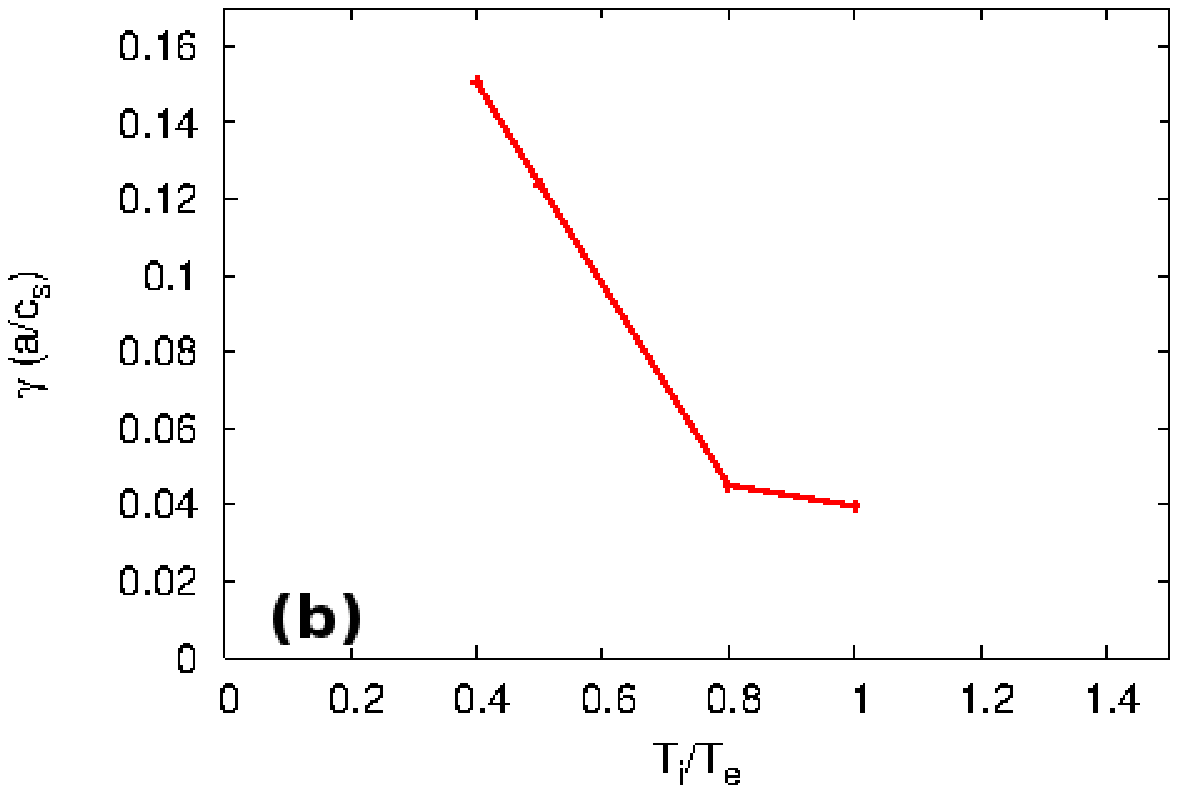}
\end{center}
\caption{(Color online) Growth rate plotted against the temperature ratio $T_i/T_e$ for $k_\theta\rho_s=1.488$. Shown are MT (a) and ITG (b).}
\label{tauscan}
\end{figure}

The MT mode is known to be driven by the electron temperature gradient (rather than the current gradient, as is the case for the large scale tearing mode). Thus, a strong dependence of the growth rate on $L_{Te}$ is expected, and this is seen in Fig.~\ref{ltscan}. Here the mode can be seen to require the threshold gradient of approximately $a/L_{Te}\approx3.5$, which is higher than that reported in \cite{Predebon2010a} ($\left(a/L_{Te}\right)_{\mathrm{crit}}\sim2$). This threshold falls at a similar value as is observed for the ion temperature gradient threshold for the ITG mode. However, it can be seen that the MT growth rate rises much more steeply with temperature. Such a strong dependence can be expected to lead to profile stiffness, fixing the experimental gradient near the threshold for instability. Additionally, nonlinear simulations in the spherical tokamak have revealed a nonlinear upshift in the effective gradient threshold as compared to linear simulations \cite{Guttenfelder2011}, and the same effect can be expected to occur in the RFP.

In Fig.~\ref{tauscan} can be seen the effect of varying the temperature ratio $T_i/T_e$. MT shows a strong increase of the growth rate with this ratio. There is an opposite dependence in ITG (see plot b), where the growth rate decreases as $T_i/T_e$ increases. This is consistent with MT's electron temperature gradient drive.

\begin{figure}[p]
\begin{center}
\includegraphics[angle=270,scale=0.6]{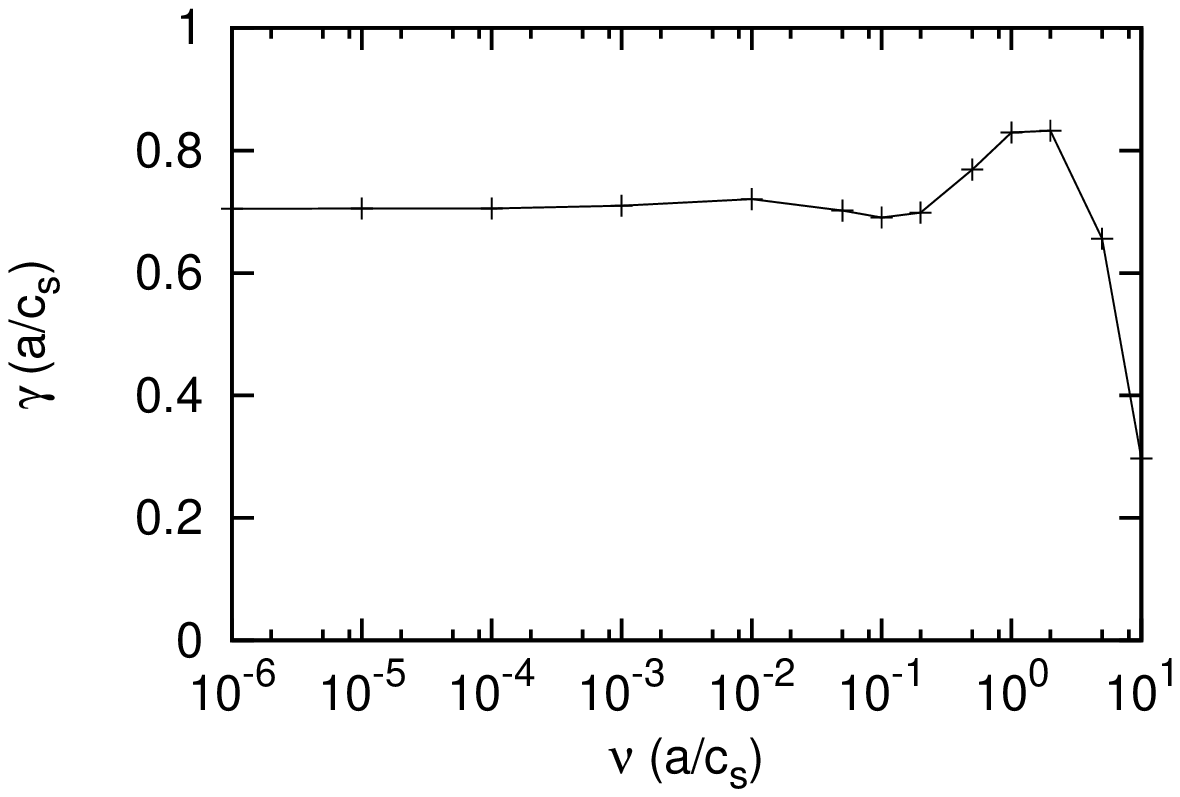}
\includegraphics[angle=270,scale=0.6]{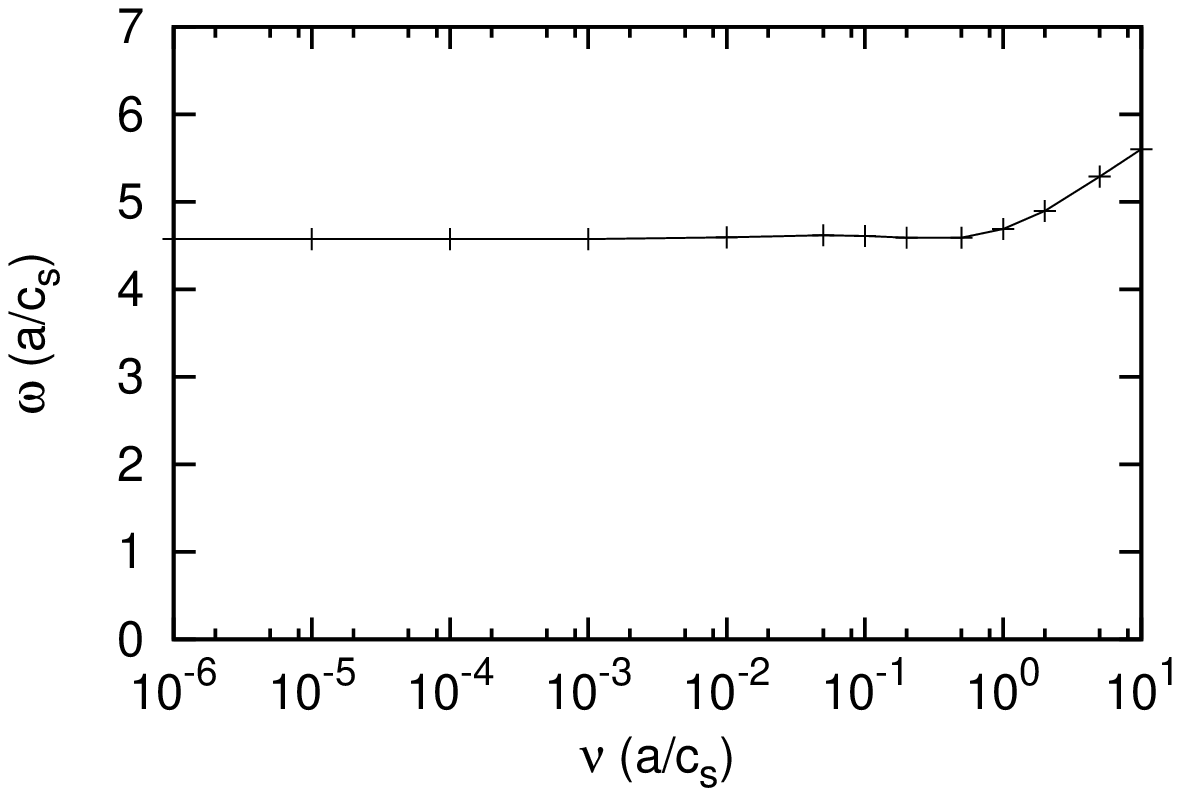}
\end{center}
\caption{MT growth rate and frequency plotted against the collisional frequency $\nu$ for $k_\theta\rho_s = 1.488$. There appear to be two distinct regmes: a region of constant growth rate and constant real frequency at low $\nu$ and a separate region at $\nu$ with a peak in growth rate and a real frequency that scales linearly with $\nu$.}
\label{collscan}
\end{figure}

Since collisions are expected to play an important role in the mechanism for instability of MT \cite{Drake1980,Connor1990}, a scan was performed over the collision frequency, the results of which are given in Fig.~\ref{collscan}. The collision operator used in GYRO is a pitch-angle scattering operator

\begin{equation}
C(f) = \frac{\nu_e(\epsilon)}{2}\frac{\partial}{\partial\xi}(1-\xi^2)\frac{\partial f}{\partial \xi},
\end{equation}
where $\xi$ is the pitch angle and $\nu_e(\epsilon) = \nu[Z_{\mathrm{eff}}+H(\epsilon^{1/2})]/\epsilon^{3/2}$, with $Z_{\mathrm{eff}}$ the effective nuclear charge, $H(x)=\mathrm{exp}[-x^2]/x\pi^{1/2}+(1-1/2x^2)\mathrm{erf}(x)$ and $\epsilon=E_e/T_e=m_e v^2/2T_e$. In these expressions, $\nu$ is the control parameter in the simulations.


A notable feature of Fig.~\ref{collscan} is the appearance of what seem to be two separate regimes of MT, with a transition between the two occuring roughly around $\nu \sim 0.1-1$. At $\nu\sim 1$ the growth rate achieves a peak and then falls off for higher collisionality. At lower values of $\nu$ the growth rate flattens, remaining finite in the limit of zero collisionality (although GYRO uses an upwind differencing scheme, which may introduce collisional effects). It should also be noted that the real frequency scales linearly with $\nu$ above $\nu \approx 1$. Previous analytic work in tokamaks (see Refs.~\cite{Drake1980,Connor1990}) has concluded that MT should be stable in the collisionless limit, though some evidence contradicting this conclusion has been seen in gyrokinetic simulations in tokamak \cite{Doerk2011} and spherical tokamak \cite{Applegate2007} geometries.

\begin{figure}[p]
\begin{center}
\includegraphics[angle=0,scale=0.5]{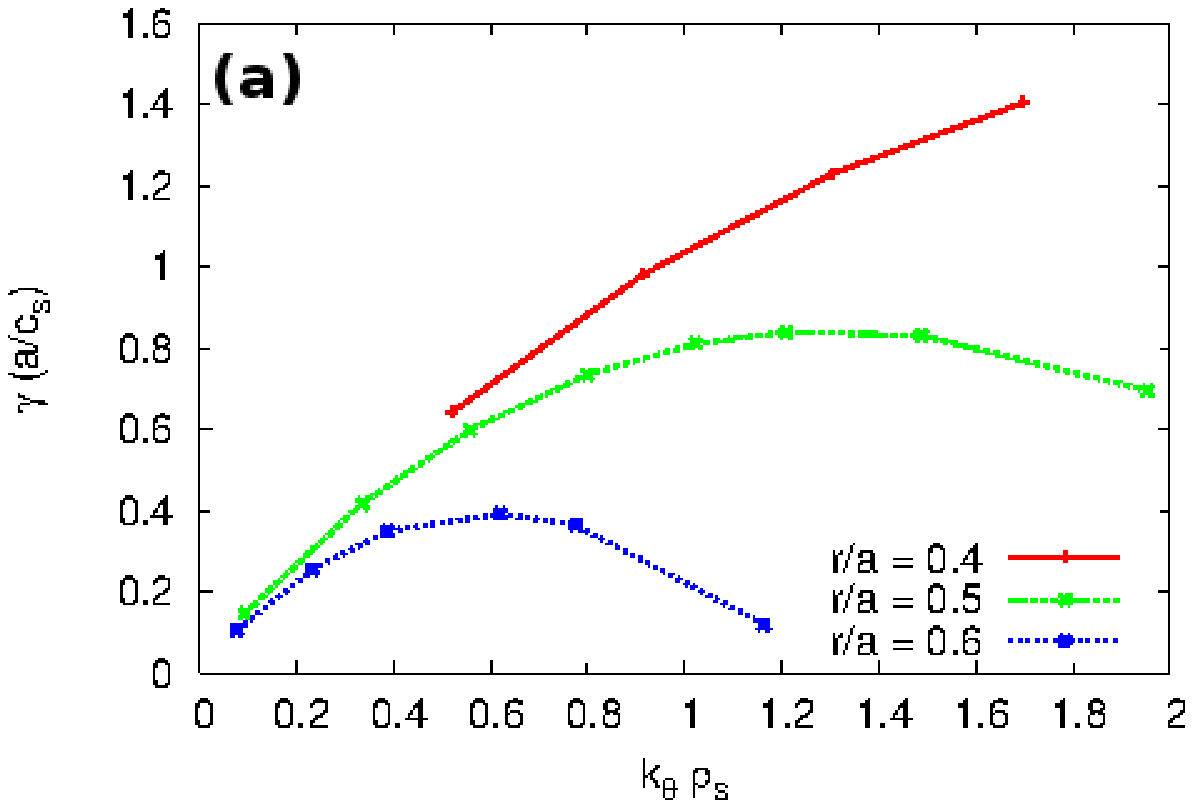}
\includegraphics[angle=0,scale=0.5]{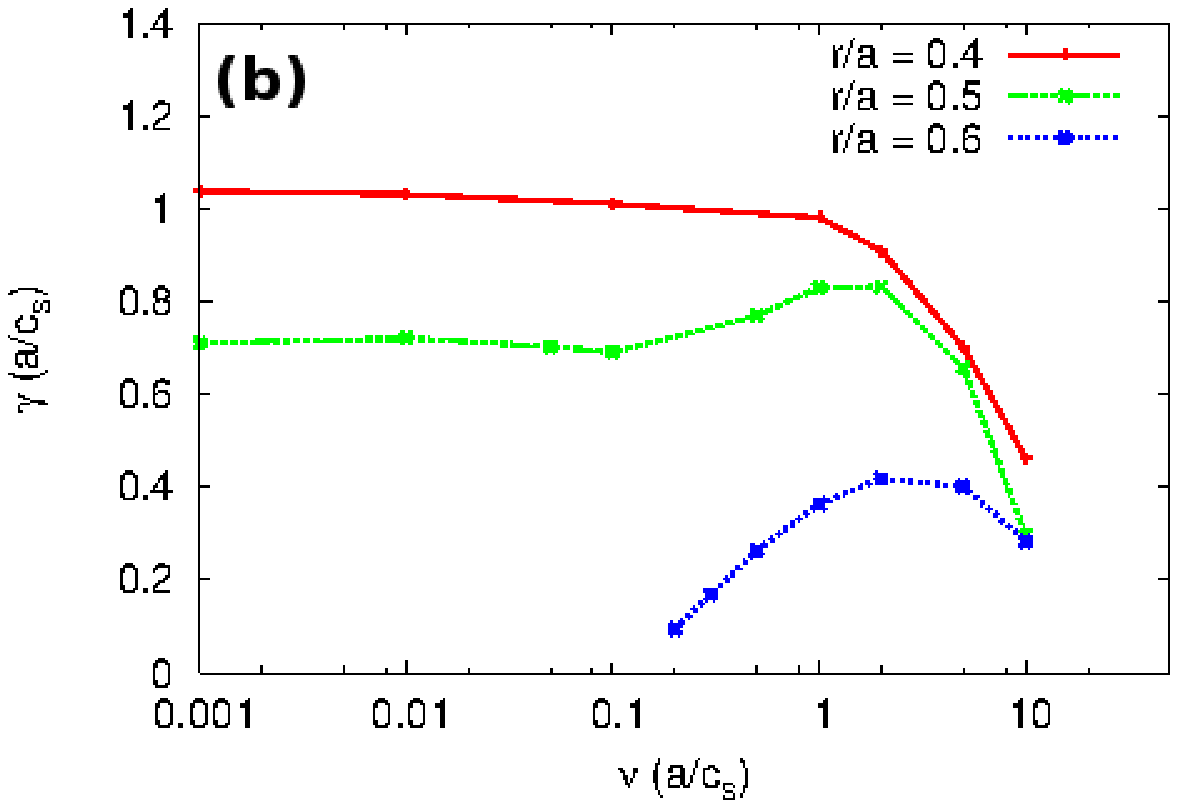}
\end{center}
\caption{(Color online) MT wavenumber spectrum (a) and  collisionality scan (b) for different values of $r/a$. The corresponding values of $q_0$ and shear are - $r/a=0.4$: $q_0=0.209$, $\hat s=-0.382$ (red solid curve); $r/a=0.5$: $q_0=0.186$, $\hat s=-0.716$ (green dashed curve); $r/a=0.6$: $q_0=0.155$, $\hat s=-1.344$ (blue dotted curve). There is stabilization of MT with increasing radius $r/a$, especially prevalent at low $\nu$. Increased radius coincides with increased shear, which may play a role in stabilization.}
\label{radial_scan}
\end{figure}

The collisional dependence of this mode was investigated at several different radii (varying only $q_0$ and $\hat s$ in correspondence with the TBFM and keeping all other parameters fixed). The results of this can be seen in Fig.~\ref{radial_scan}. Importantly, the wavenumber spectrum (a) shows a general stabilization of the mode at larger radius, which might be attributed to the larger shear at these locations. The collisionality scan (b) also shows interesting behavior: in particular, the growth rate at low $\nu$ is much more affected by increased radius than at $\nu\sim 1$, and at $r/a=0.6$ the mode is completely stabilized in the collisionless limit. This behaviour may suggest that there are two distinct varieties of MT modes in these simulations -- one at low collisionality and one at moderate collisionality -- that have somewhat different physics behind their drive mechanisms. The effects of increasing the pinch parameter $\Theta$ are similar to the effects of increasing the radius. This is to be expected from the TBFM, due to the same parametric dependence on $r$ and $\Theta$ in that model.

\begin{figure}[p]
\begin{center}
\includegraphics[angle=0,scale=0.5]{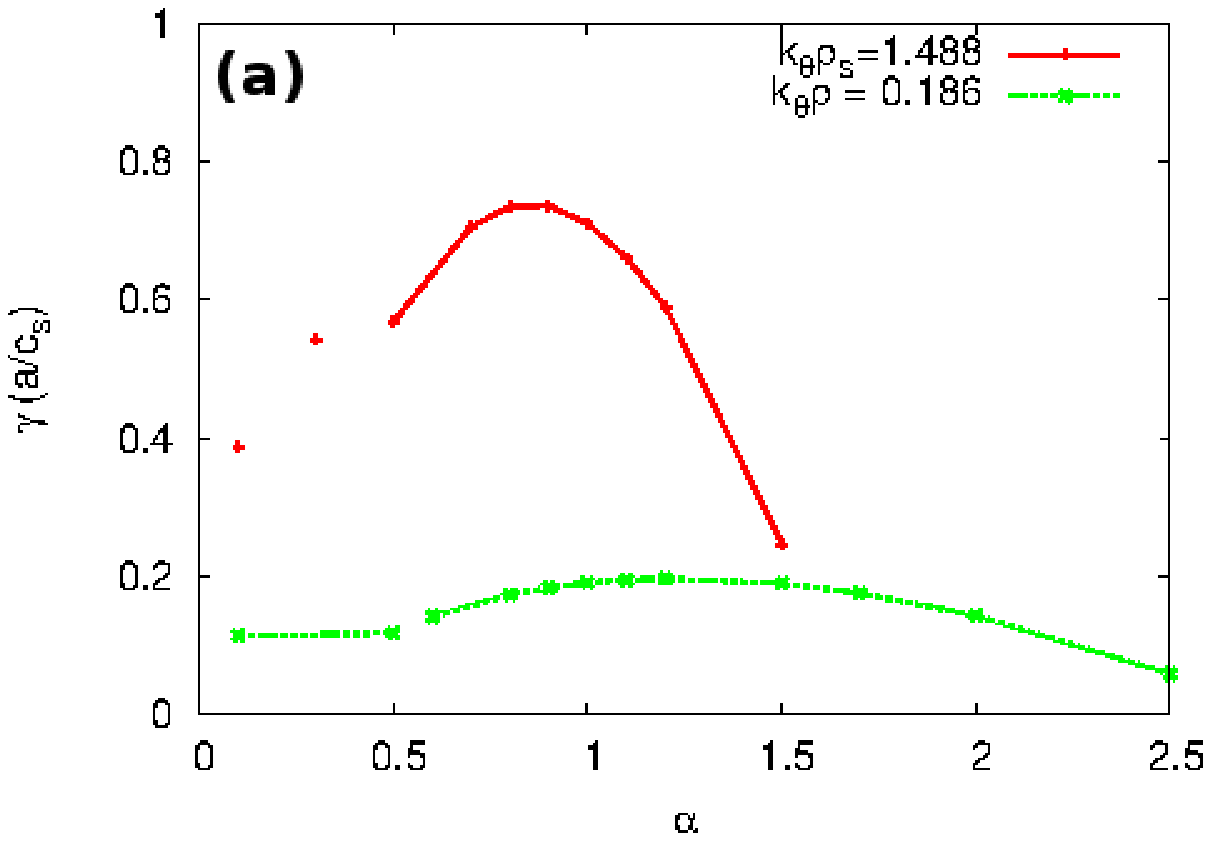}
\includegraphics[angle=0,scale=0.5]{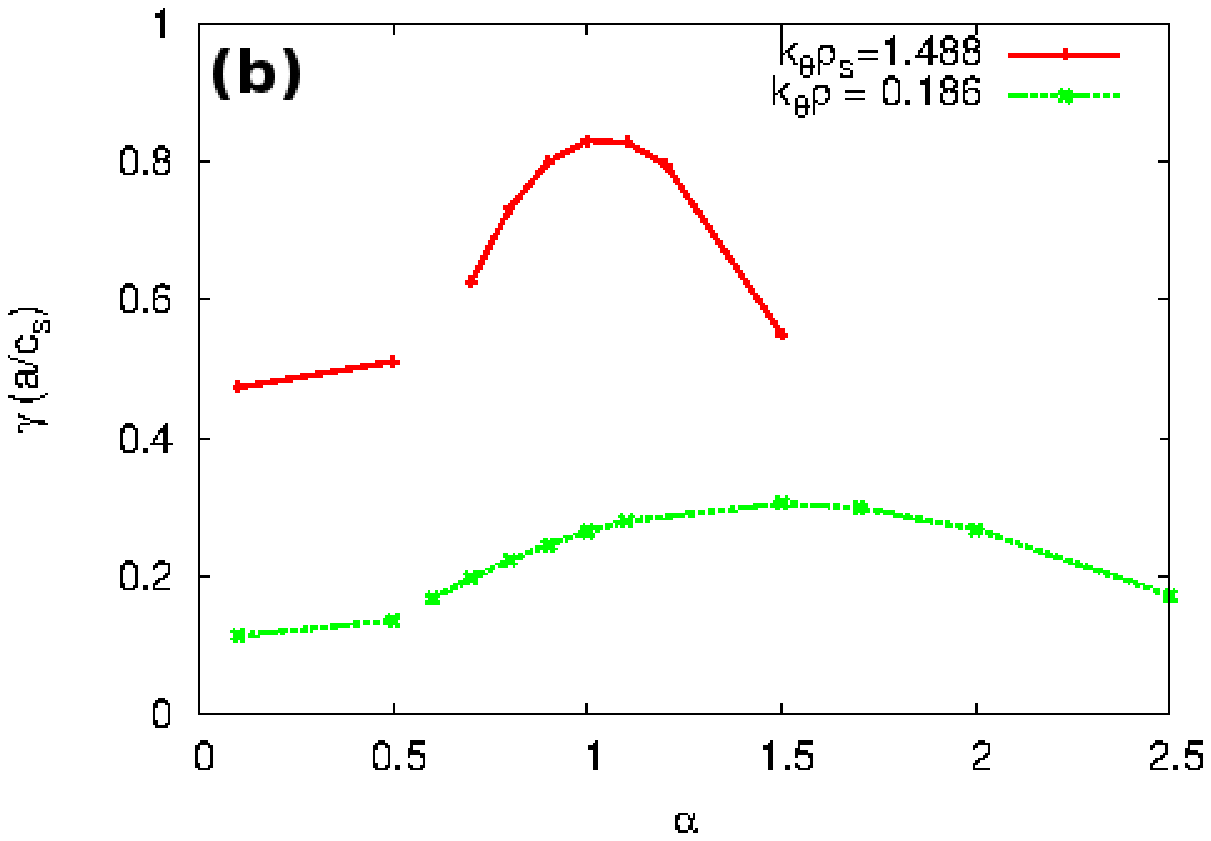}
\end{center}
\caption{(Color online) The role of curvature drift in the MT instability. The parameter $\alpha$ is a factor regulating the strength of the curvature drift in the code. Shown are $\nu=0.001$ (a) and $\nu=1.0$ (b), as well as $k_\theta \rho_s = 1.488$ (red solid curve) and $k_\theta \rho_s = 0.186$. This behavior is similar to that seen in Ref.~\cite{Finn1986}. The points at low $\alpha$ represent a separate mode that has not been studied in detail.}
\label{curvatureDrift}
\end{figure}

The physics of the drive mechanism was investigated further by looking at the role of curvature drift in the instability. In this study, a scalar factor $\alpha$ was placed in front of the curvature drift term in GYRO and varied to change its relative strength; a value of $\alpha = 1$ corresponds to the physical curvature drift. The results of this are seen in Fig.~\ref{curvatureDrift} for $\nu=0.001$ (a) and $\nu=1.0$ (b). In both cases, the instability is strongest for $\alpha\approx 1$ and falls off for values much lower or higher than this. This behavior is consistent with the magnetic curvature drift instability derived by Finn and Drake (see Ref.~\cite{Finn1986}), in which they describe a semi-collisional drift-tearing mode unstable in the presence of curvature drift and an electron temperature gradient. Ref. \cite{Finn1986} is formulated in a cylindrical RFP equilibrium, and they note that this instability is strong only when $\omega_D \sim \omega_{*}$, a characteristic that makes it more relevant to the RFP, with its associated stronger curvature drifts, than in a standard tokamak. Spherical tokamaks might also be expected to be susceptible to the curvature drift instability. We also note that although Ref. \cite{Finn1986} uses fluid theory in the semi-collisional regime, collisions do not play an explicit role in the instability, and it is plausible that this mode may arise in the collisionless limit with proper inclusion of curvature drifts.

\section{Conclusion \label{conc}}

The linear characteristics of microinstabilities in RFP geometry were investigated using the gyrokinetic code GYRO. It was found that ITG is the dominant gyroscale instability at low $\beta$ for the parameters used in this work, with an increase in $\beta$ causing a transition to MT. An analysis of the finite-$\beta$ suppression of ITG demonstrated both qualitative and quanititative agreement with the simulation results. This analysis suggests that the suppression mechanism in the RFP is the same as in the tokamak, with a larger critical $\beta$ that arises predominantly from a shorter parallel connection length.

The MT mode was further investigated by performing a variety of parameter scans. The characteristics of the mode were in agreement with previous simulation work in other devices. Evidence for a collisionless MT mode was seen, though further work remains to be done as to be determine the nature of the physics behind this mode. There is some evidence that curvature drift plays an important role in the instability, and it may be described by the magnetic curvature drift instability of Finn and Drake \cite{Finn1986}.

The results of this work regarding implications for the Madison Symmetric Torus suggest that MT may indeed be playing a role in the gyroscale dynamics. Standard discharges lie in the range of $\beta$ for which both ITG and MT may be excited, while PPCD discharges achieve $\beta$ values high enough so that ITG may be sufficiently suppressed and only MT is important. For these conditions and a large enough temperature gradient, the strength of MT and range of scales at which it exists suggest that it should be making important contributions to gyroscale turbulence. Let us note, however, that for alternate sets of parameters other modes, such as the trapped electron mode or kinetic ballooning mode, may emerge and play a role in gyroscale dynamics. Further work must be completed to determine the critical parameters required for these various instabilities and the effect of the background equilibria on their growth rates and transport fluxes. This task, as well as measuring and characterizing these modes experimentally, is an important area of future work.

\section*{Acknowledgements}
The authors would like to acknowledge V.~Tangri for helpful discussions on the RFP geometry and its implemenation in GYRO, and to C.~Hegna for insights on tearing modes. This work was supported by (US)DOE Grant No.~DE-FG-02-85ER53212 and supported in part by the National Science Foundation through XSEDE resources, Grant No.~PHY120009.


\end{document}